\documentclass{cmspaper}
\usepackage{epsfig} 

\begin{document}

\renewcommand{\baselinestretch}{0.9}
\newcommand{\ccaption}[2]{
    \begin{center}
    \parbox{0.85\textwidth}{
      \caption[#1]{\small{{#2}}}
      }
    \end{center}
    }

\newcommand     \HP             {{$H^{\pm}$}} 
\newcommand     \ETM  {{E_{\top} \hspace{-14pt} \backslash \hspace{8pt}}} 
\newcommand     \TTF {\ttfamily}


\setlength{\fboxrule}{1pt}
\setlength{\fboxsep}{3mm}
\newcommand{\drawbox}[1]{\vspace{\baselineskip}\noindent%
\fbox{\texttt{#1}}\vspace{0.5\baselineskip}}
\newcommand{\itemc}[1]{\item[\textbf{#1}\hfill]}
\newcommand{\iteme}[1]{\item[\texttt{#1}\hfill]}
\newcommand{\itemn}[1]{\item[{#1}\hfill]}

\newenvironment{entry}%
{\begin{list}{}{\setlength{\topsep}{0mm} \setlength{\itemsep}{0mm}
\setlength{\parskip}{0mm} \setlength{\parsep}{0mm}
\setlength{\leftmargin}{20mm} \setlength{\rightmargin}{0mm}
\setlength{\labelwidth}{18mm} \setlength{\labelsep}{2mm}}}%
{\end{list}}%
\newenvironment{subentry}%
{\begin{list}{}{\setlength{\topsep}{0mm} \setlength{\itemsep}{0mm}
\setlength{\parskip}{0mm} \setlength{\parsep}{0mm}
\setlength{\leftmargin}{10mm} \setlength{\rightmargin}{0mm}
\setlength{\labelwidth}{18mm} \setlength{\labelsep}{2mm}}}%
{\end{list}}


\begin{titlepage}
   \cmsnote{2002/010}
   \date{March 8, 2002} 

 \title{\Large \sf Study of $s$-channel Charged Higgs Production in CMS }

  \begin{Authlist}
S.R.~Slabospitsky\Aref{*}
\Instfoot{ihep}{Institute for High Energy Physics, \\ Protvino, Moscow Region,
 RUSSIA } 
  \end{Authlist}

\Anotfoot{*}{\tt Serguei.Slabospitski@cern.ch } 

\begin{abstract} 
CMS potential for study of the $s$-channel \HP-boson production via light 
quark annihilation is investigated for large
values of $\tan\beta$ ($=20\div 50$) and relatively light charge Higgs 
boson ($M_H = 200 \div 400$~GeV). 
An appropriate parameterisation for $M_{\top}(j, \ETM)$-distributions
of the signal and background events is proposed, which provides the
determination of the charged Higgs mass and $\tan\beta$ parameter.
\end{abstract}                                                          

\vskip 1cm
\vfill 

\end{titlepage}

\setcounter{page}{2}

\section*{\bf Introduction }
   
A charged Higgs boson (\HP) appears in many well motivated extensions of the
Standard Model (SM) (i.g. in the MSSM~\cite{Nilles:1984ge}).  
Two free  parameters, $M_H$ and $\tan\beta$, determine all properties and
couplings of the \HP-boson. The searches
for this object were performed in many experiments. LEP collaborations have
set lower limits in a model independent way on the mass of \HP-boson,
$M(H^{\pm}) > 78.5$~GeV for any $\tan \beta$~\cite{Holzner:2001tv}. 
Two experiments
at the Tevatron, CDF and D$\emptyset$, have performed several searches for
\HP~\cite{Abe:1997rk}. They excluded the low $(<1)$ and high 
$(>40)$ $\tan\beta$
region up to 120~GeV and 160~GeV, respectively. The additional indirect limit
can be evaluated from low-energy physics, for example from the values of
$D_s \to \tau \nu$ and $B \to \tau \nu$ decay branching 
ratios~\cite{Mangano:1997md}. 

The discovery potential of \HP at LHC has been investigated by both 
ATLAS~\cite{atlas} and CMS~\cite{cms, Denegri:2001pn} collaborations. 
It was established that 
for heavy charged Higgs with $M(H^{\pm}) > m_t$ the most promising channel is
top-Higgs associated production in two subprocesses~\cite{roy}
\[
g b \to t H^{\pm}, \quad g g \to t \, H^{\pm} \, \bar b
\]

In the present work we consider an additional subprocess of \HP~ production 
due to  annihilation of the light $q \bar q'$-pair from
the initial hadrons (protons): 
\[
 q \, \bar q' \, \to \, H^{\pm}, \quad q = d, u, s, c, b  
\] 
The \HP-boson production in this $s$-channel with subsequent 
\HP decay into $t \bar b$ pair was investigated earlier
(see~\cite{tait, Foursa:2000yu, Diaz-Cruz:2001gf}).
Here we investigate the CMS potential for study of the $s$-channel \HP-boson 
production via light quark annihilation with subsequent
\HP decay into a $\tau^{\pm} \nu_{\tau}$ lepton pair
(see also~\cite{Diaz-Cruz:2001gf}):
%
%
%
\[
q \bar q' \to  H^{\pm} \to \tau^{\pm} \nu_{\tau}
\]
We perform our calculation for the case of $pp$-collisions at
$\sqrt{s} = 14$~TeV:
\begin{equation}
 p p \to H^{\pm} X \label{pp}
\end{equation}
for large values of $\tan\beta$ and for four values
of charged Higgs mass:
\begin{equation}
 \tan\beta = 50 \quad {\rm and} \quad  M_H = 200, \, 250, \, 300, \, 400\,\,
{\rm GeV} \label{defpar}
\end{equation}
We use CTEQ5L parameterisation~\cite{Lai:1999wy} of the parton distributions. 
All
estimates of the expected number of events are performed for three year 
low-luminosity run of LHC:
\begin{equation}
  L_{int} \equiv \int {\mathcal L} dt = 30 \,\,{\rm fb}^{-1} \label{lumin}
\end{equation}

For calculation of the signal and background processes we use 
the event generator {\sf TopReX}~3.25~\cite{toprex}, while the well-known
package PYTHIA~6.157~\cite{pythia} is explored for modelling of quark and 
gluon
hadronisation. For proper simulation of the detector response all generated
events are passed through the fast Monte Carlo package 
CMSJET~4.703~\cite{cmsjet}.

We find that even after application of all appropriate cuts
the expected number of the signal events is relatively large
($\sim 10^2 \div 10^3$) for $M_H = 200 \div 400$~GeV and 
$\tan \beta \sim 40 \div 50$. Therefore, this $s$-channel  \HP-boson 
production process allows not only to establish the presence of the signal 
from \HP-boson,
but also makes possible to measure the parameters (the mass and $\tan\beta$) of
the charged Higgs boson.

\section{\bf Signal event generation }

\begin{figure}[t] 
\begin{center}
\epsfig{file=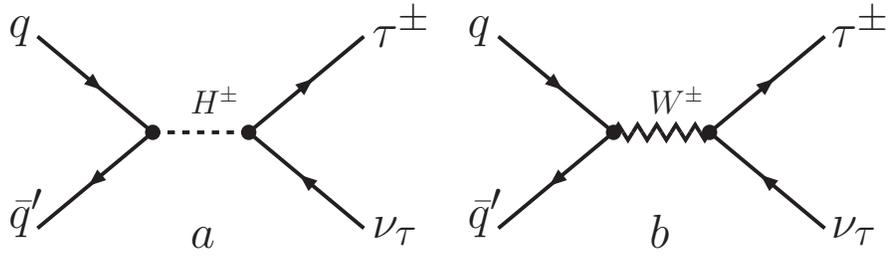,width=12cm,clip=} 
\caption{Diagrams describing light quarks ($q = d, u, s, c, b$)
annihilation into $\tau^{\pm} \nu_{\tau}$ pair via charged \HP-boson~(a) and
$W^{\pm}$-boson~(b) exchange. } \label{fig1} 
\end{center}
\end{figure}

Fig.~\ref{fig1}a presents the diagram describing the process under 
consideration,
\begin{equation}
 q \, \bar q' \, \to \, H^{\pm}, \quad q = d, u, s, c, b \label{reac1}
\end{equation}
 
Note, that Higgs boson couplings to fermions are proportional to the masses of
these fermions~\cite{Nilles:1984ge}. Therefore, the corresponding production
cross section has strong dependence on the light quark mass values. 
In our calculations we use so-called 
``current'' values of $m_q$~\cite{Groom:2000in}:
\begin{eqnarray*}
\begin{array}{lll}
 m_d = 0.009\,\,{\rm GeV,} & m_u = 0.005\,\,{\rm GeV,} &
 m_s = 0.150\,\,{\rm GeV,}  \\
  m_c = 1.250\,\,{\rm GeV,} &  m_b = 4.800\,\,{\rm GeV,} &
\end{array}
\end{eqnarray*}
which are smaller (especially for light $d, u$, and $s$ quarks)
then those values of quark masses  used in PYTHIA ($m_d=m_u=300$~MeV, 
etc, see~\cite{pythia}).

\begin{figure}[tb]
\vspace{3.truecm}

\begin{center}
\epsfig{file=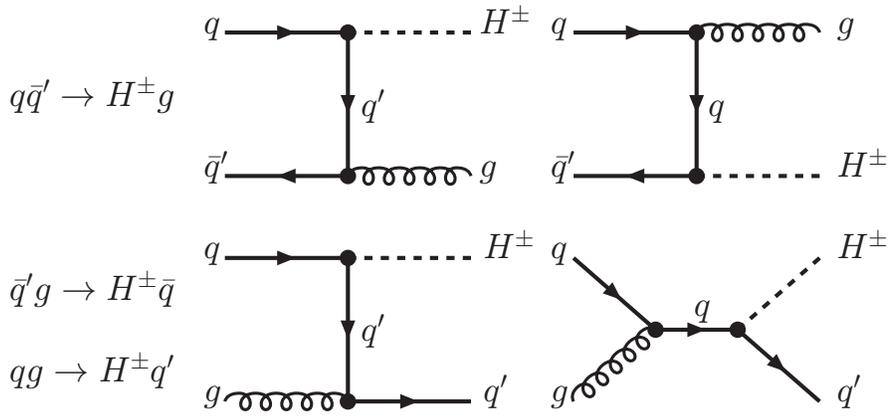,width=12cm,clip=} 
\caption{Diagrams describing NLO corrections to $2 \to 1$ process of
\HP-boson production ($q = d, u, s, c, b$) } \label{fig2} 
\end{center}
\end{figure}

We also take into account the radiative QCD corrections to the $2 \to 1$ 
process~(\ref{reac1}). In so doing 
we calculate also the NLO processes ($2 \to 2$) (see diagrams in 
Fig.~\ref{fig2})
\begin{equation}
 q \, \bar q' \, \to \, H^{\pm} \, g, \quad
 q \, g \, \to \, H^{\pm} \, q', \quad
 \bar q' \, g \, \to \, H^{\pm} \, \bar q \label{reac2}
\end{equation}

It is well-known that the consideration of such processes at small
$\hat k_{\bot}$ (where  $\hat k_{\top}$ is the transverse
momentum of the final particle, \HP, $q$ or $g$, defined in the 
centre-of-mass system of the scattering partons) leads to a double-counting 
problem. Indeed, in this region ($\hat k_{\bot} \to 0$)
the virtual quark, entering the $q \bar q' H^{\pm}$ vertex, has very 
small virtuality and may be considered as a on-shell parton. As a 
result, any $2 \to 2$ process~(\ref{reac2}) can be factorized into two 
subprocesses. The first one is the initial parton ($q$ or $g$) splitting 
into two 
partons
\[ q \to q g, \,\, \bar q' \to \bar q' g, \,\, g \to q \bar q.
\] 
The second subprocess is the quark-antiquark annihilation into \HP, where
one quark (antiquark) comes from initial hadron, while the second quark 
appears due to parton splitting.  However,  such a process was also calculated 
early as the process~(\ref{reac1}).

This problem was considered in details in~\cite{Balazs:1998sb}, where
the calculations of the complete ${\cal O}(\alpha_s)$ QCD corrections 
(reactions~(\ref{reac2})) to the $s$-channel production process~(\ref{reac1}),
including QCD-resummation over multiple soft-gluon emission
was performed. 
In our numerical calculations we use an approximation, which
provides desirable accuracy (see~\cite{boos} 
for details). It is based on the consideration of the distribution on
the charged Higgs transverse momentum, $p_{\top}(H)$, defined in the initial 
$pp$ reference frame. In the region of small $p_{\top}(H)$ the basic 
contribution to Higgs production comes from $2 \to 1$ process~(\ref{reac1}),
 while the $2 \to 2$ process~(\ref{reac2}) is responsible
 for Higgs production with high $p_{\top}(H)$.

The method of event generation is thus as follows. Firstly, we generate
 events with \HP-boson production due to 
 $2 \to 1$ process~(\ref{reac1}). Any event from this sample will be accepted 
if the transverse momentum of the charged Higgs is smaller than some value  
$p_0$. Then we generate the second sample of events due to $2 \to 2$ 
process~(\ref{reac2}) with final parton transverse
momentum $\hat k_{\top} > \hat k_0$. Any event from this second sample will be 
accepted if  $p_{\top}(H) > p_0$. Thus, the resulting (total) sample of 
\HP-boson production events is the sum of two contributions:
\begin{eqnarray*}
N(p p \to H^{\pm} X) &=& N^{(2\to1)}(p p \to H^{\pm}; \,
p_{\top}(H) < p_0) \\
&+& N^{(2\to 2)}(p p \to H^{\pm}\, jet;  \hat k_{\top} > \hat k_0,
p_{\top}(H) \geq p_0) 
\end{eqnarray*}

\begin{figure}[t]
\begin{center}
\epsfig{file=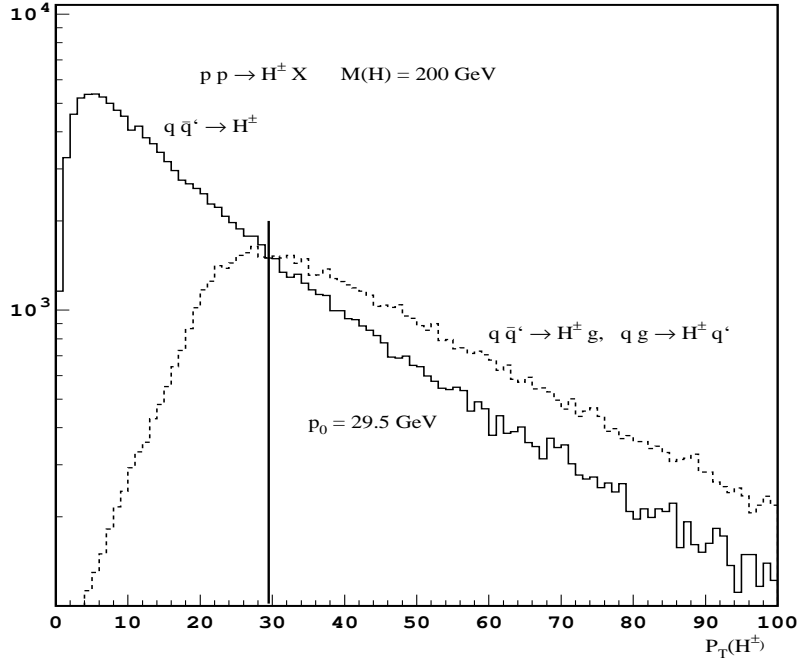,width=12cm,height=10cm,clip=} 
\caption{ $p_{\top}(H)$-distribution of \HP-boson production in LO
subprocess~(\ref{reac1}) (solid histogram) and in NLO 
subprocess~(\ref{reac2}) (dashed histogram). The vertical line corresponds
to parameter value  $p_0 = 29.5$~GeV (see text). }  \label{fig3} 
\end{center}
\end{figure}

We find that the  smooth behaviour of the resulting 
$p_{\top}(H)$-distribution can be achieved for the following values of 
these parameters:
\begin{eqnarray*}
     && \hat k_0 \approx 20\,\,{\rm GeV,} \\
 {\rm and} && p_0 = 29.5 \,\,{\rm GeV,}\,\,{\rm for} \,\,
M_H = (200 \div 400) \,\,{\rm GeV}
\end{eqnarray*}
The corresponding $p_{\top}(H)$-distributions are shown in 
Fig.~\ref{fig3}.

\begin{figure}[b]
\begin{center}
\epsfig{file=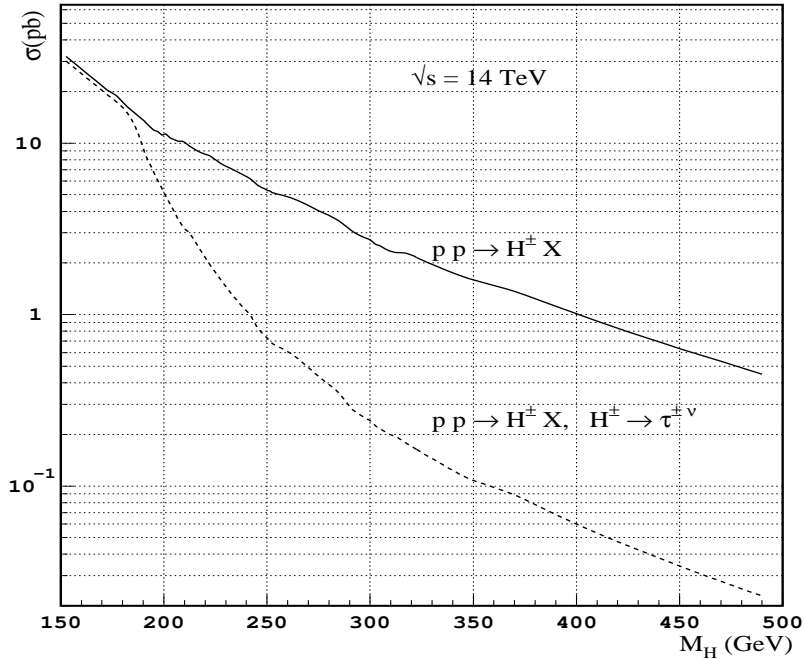,width=12cm,height=10cm,clip=} 
\caption{Behaviour of the total production cross section  for
\HP-boson in $pp$-collisions (reaction~(\ref{pp})) at 
$\sqrt{s} = 14$~TeV and $\tan\beta=50$ versus charged Higgs boson mass, 
$M_H$ (solid curve). The dashed curve represents the same cross section times 
branching fraction to $H^{\pm} \to \tau^{\pm} \nu_{\tau}$. } \label{fig4}
\end{center}
\end{figure}

\begin{figure}[htb]
\begin{center}
\epsfig{file=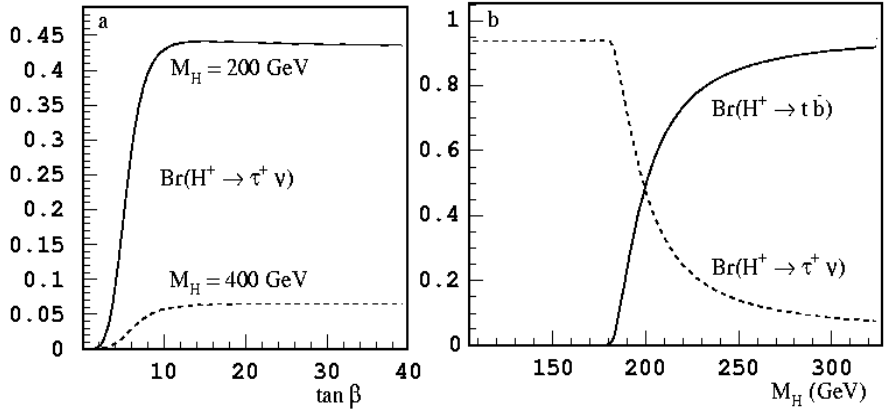,width=12cm,height=6cm,clip=} 
\caption{ The behaviour of the branching fractions of \HP-boson decays into
$\tau^{\pm} \nu_{\tau}$ and $t \bar b$ pairs versus $\tan\beta$~$(a)$ and
$M_H$~$(b)$. } \label{fig5} 
\end{center}
\end{figure}

The behaviour of the production cross section  for $\tan\beta = 50$ versus
$M_H$ is shown in Fig.~\ref{fig4}. For $\tan\beta \geq 10$ the 
branching ratio of the $H^{\pm} \to \tau^{\pm} \nu_{\tau}$ decay is almost 
independent on $\tan\beta$ (see Fig.~\ref{fig5}a). Therefore, the 
$\tan\beta$ dependence of the production cross section for \HP has a very 
simple quadratic dependence:
\begin{equation}
\sigma(p p \, \to \, H^{\pm} X) \propto \tan^2\beta  \label{pp2}
\end{equation}
At the same time the Br$(H^{\pm} \to \tau^{\pm} \nu)$ has a strong dependence
on the mass of charged Higgs in the region of $M_H = 200\div400$~GeV (see 
Fig.~\ref{fig5}b) due to opening of the $H^{\pm} \to t \bar b$ decay
channel. Therefore, a simultaneous measurement of 
the charged Higgs production cross section and its mass provides a possibility 
for indirect determination of the $\tan\beta$ parameter value.

\section {\bf Signal/background separation }

The most important and irreducible background comes from 
$\tau^{\pm} \nu_{\tau}$ production via virtual (Drell-Yan) $W^{\pm}$-boson 
exchange 
(see Fig.~\ref{fig1}b). All other possible sources of background give 
relatively small contributions and will be not considered.

It is well-known that due to the different nature of $H^{\pm} f \bar f'$ and
$W^{\pm} f \bar f'$ interactions the final $\tau$-leptons, produced via 
$H$ or $W$ boson exchanges have opposite polarisations (see~\cite{roy}).
This feature provides an effective way for background suppression.
In particular, we use the hadronic $\tau \to \pi \nu$ decay mode, where
the the two opposite $\tau$ helicity states lead to remarkably different
decay pion laboratory momentum distributions.

\begin{figure}[htb]
\begin{center}
\epsfig{file=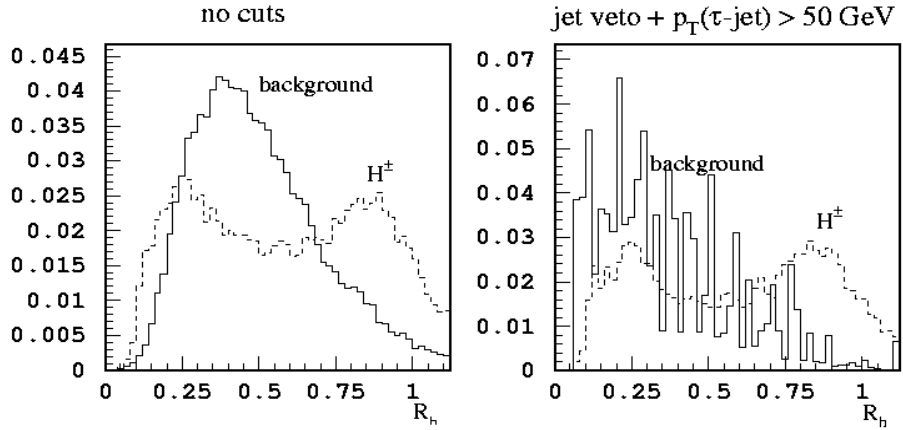,width=12cm,height=6cm,clip=} 
\caption{ The number of events distribution versus 
 $R_h \equiv E(h^{\pm}) / E({\rm jet}$. The solid (dashed)
histograms correspond to the background (signal) events. The left histogram
is  before  any cuts, while the right one
 corresponds to requirements of ``jet-veto'' and  
$E_{\top}(\tau$-jet),~$\ETM) \ge 50$~GeV. }  \label{fig6} 
\end{center}
\end{figure}

For hadronic $\tau$ decay identification and reconstruction we use algorithm 
which allows to identify $\tau$-jets in the one-prong decay mode 
(see~\cite{cms} for details). This algorithm is based on the fact that the 
hadronic decays of $\tau$-leptons from $H^{\pm} \to \tau^{\pm} \nu$ decay 
are seen as a narrow, low multiplicity  ``jet'' 
with a large fraction of calorimetric energy consisting from a single track. 
Due to the opposite polarisation of $\tau$-leptons produced via 
$H$/$W$ boson decays the fraction of the total $\tau$-jet energy carried 
away by the charged track relative to the parent $\tau$ energy is very
different. 
For a $W$-boson mediated  decay such a track carries away a significantly 
smaller fraction of the $\tau$ energy then for the \HP decay. It can be seen 
in Fig.~\ref{fig6}, which shows the signal (\HP) and background ($W^*$) event 
distribution for $R_h$ variable:
\[  R_h \equiv E(h^{\pm}) / E({\rm jet})
\]
The best signal-to-background separation in terms of this variable is
achieved for $R_0 \geq 0.8$. 

To suppress the large $W$~$+$~jet(s) background with $W \to \tau \nu$
($\tau \to$~hadron~$+\,\,\nu$) decays we also require a strong central jet 
veto: no other jet with $E_{\top} \geq 20$~GeV in $|\eta| \le 4.5$ 

To attempt to extract a $H^{\pm} \to \tau^{\pm} \nu$ 
(with $\tau^{\pm} \to \pi^{\pm}\, + \, X^0 \, + \, \nu$) signal we thus 
require: 
\begin{itemize}
\item[$\bullet$] one identified $\tau$-jet with $E_{\top} \ge 50$~GeV and
 $|\eta| < 2.0$
\item[$\bullet$] missing transverse energy $\ETM \ge 50$~GeV
\item[$\bullet$] jet veto: no other hadronic jets with
 $E_{\top}(j) \ge 20$~GeV in $|\eta| \le 4.5$
\item[$\bullet$] no other identified objects (leptons, photons) with 
 $E_{\top} \geq 10$~GeV in $|\eta| \le 2.4$
\item[$\bullet$] $R_h \ge R_0 = 0.8$ (to favour \HP onto $W^*$ as a source of
 $\tau$'s)
\end{itemize}

 The expected number of events for $30$~fb$^{-1}$ and corresponding signal
significances  ($N_S/\sqrt{N_S+N_B}$) after 
application of all cuts are given in Table~\ref{tab1}.


\section{\bf Parameterisation of $M_{\top}$-distribution}

Since the expected number of the signal and background events is relatively 
large, $N_{H} \sim (10^2 \div 10^3)$
and $N_{B} \sim 10^3$ (Table~\ref{tab1}), we can try to determine 
the parameters of 
\HP-boson, namely the mass of charged Higgs ($M_H$) and $\tan\beta$.
This could be done by fitting the distribution of the  transverse mass
$M_{\top}(j, \ETM)$~\footnote{In what follows the symbol ``j'' stands for
 the $\tau$-jet} 
 of the $\tau$-jet  and the missing energy:
\begin{equation}
M_{\top}^2 = (E_{\top}(j) + \ETM)^2 - (\vec p_{\top}(j) +\vec \ETM )^2,
\end{equation}
here $E_{\top}(j) (\vec p_{\top}(j))$ is the transverse energy (momentum) of
the $\tau$-jet  and $\ETM$~ is the missing transverse energy in the event.

The $\tau$-jet identification and reconstruction algorithm does not
provide reconstruction of the full $\tau$-lepton momentum due to the 
undetected 
neutrino from $\tau^{\pm} \to h^{\pm} (h^0) \nu_{\tau}$ decays. As a result, 
the well-known sharp two-body decay Jacobian peak in 
$m_{\top}(\tau, \nu)$-distribution 
transforms into a wide bump (see~\cite{cms, roy} and Fig.~\ref{fig8}) in the
observable $M_{\top}(j, \ETM)$.

 The form of this curve results from the ``convolution'' of the
theoretical $m_{\top}(\tau, \nu)$-distribution (where $\tau$ and $\nu$ are
produced in processes~(\ref{reac1} - \ref{reac2})) and the
``fragmentation'' (or ``decay'') of the produced $\tau$-lepton into the
observable hadronic $\tau$-jet.
This ``fragmentation'' 
depends on the experimental device
(detector acceptance, resolution, efficiency, etc) as well as on the $\tau$-jet
reconstruction algorithm and could not be calculated theoretically. At the 
same time, any appropriate functional form 
describing this ``fragmentation'' will provide a suitable 
parameterisation in our case.

\begin{figure}[htb]
\begin{center}
\epsfig{file=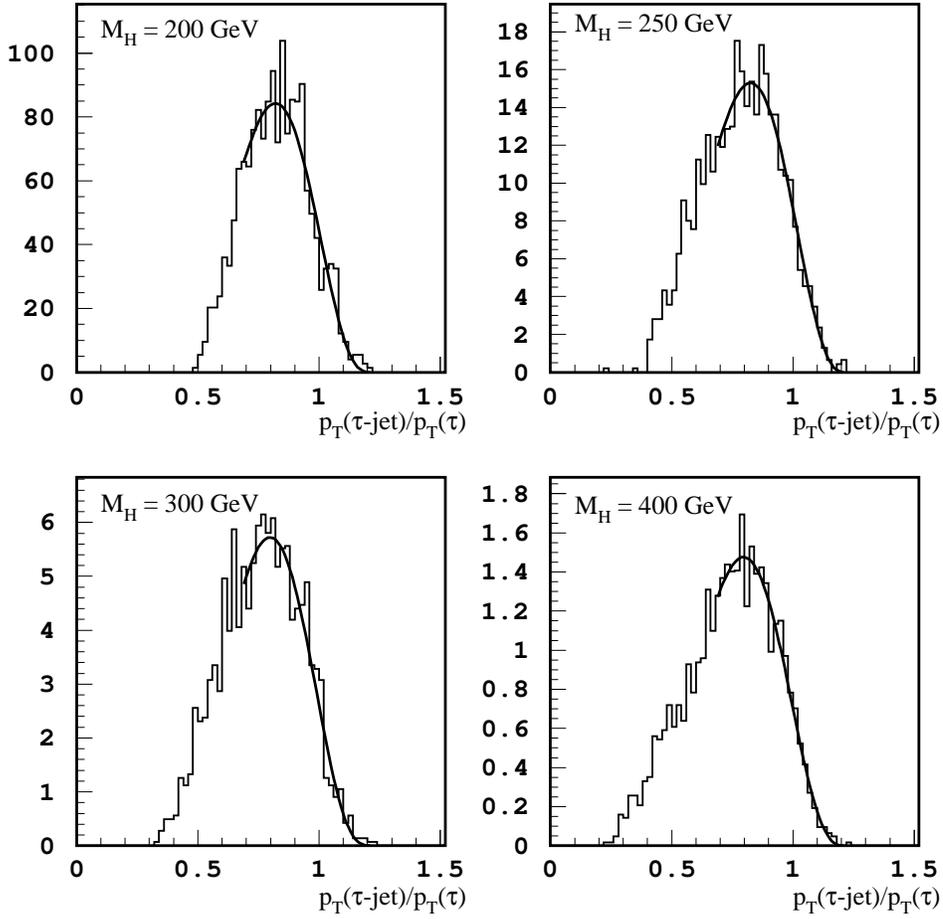,width=14cm,clip=} 
\caption{ Signal events distribution as a function of the ratio of the 
transverse momentum 
of $\tau$-jet ($p_{\top}(\tau$-jet)) to transverse momentum of parent 
$\tau$-lepton ($p_{\top}(\tau)$). The curves are results of the fit to 
``fragmentation'' function $D_{\tau \, \to \, j} (z)$ from~(\ref{dz}). }
  \label{fig7}  
\end{center}
\end{figure}

\begin{figure}[t]
\begin{center}
\epsfig{file=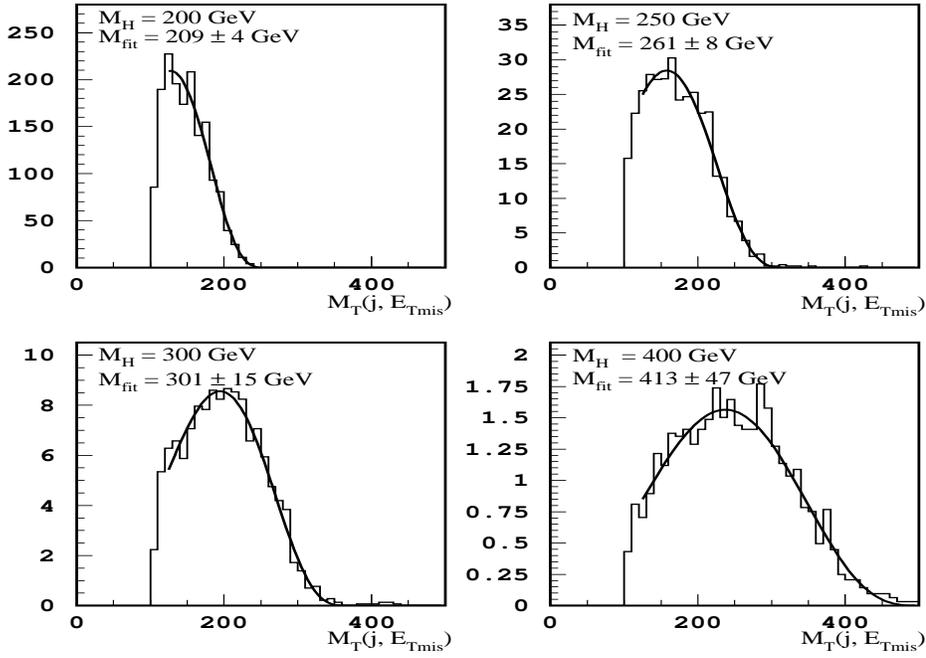,width=14cm,height=10cm,clip=} 
\caption{ $M_{\top}(j, \ETM)$-distribution for the signal events
for four input values of \HP-boson mass. The curves are results of the fit
to $F_S(M_{\top}, M_f)$ parameterisation from~(\ref{mjmis}). }
 \label{fig8} 
\end{center}
\end{figure}

For this ``fragmentation'' function $D_{\tau \to j}(z)$ we 
use a simple parameterisation as follows~\cite{klp}:
\begin{equation}
   D_{\tau \, \to \, j} (z) \propto z^{\alpha} \, (z_0 - z)^{\lambda},
\label{dz}
\end{equation}
where the scaling variable $z = {p_{\top}(j)} /{p_{\top}(\tau)}$ is the ratio 
of the transverse momentum of the $\tau$-jet (${p_{\top}(j)})$ to
the transverse momentum of the parent 
$\tau$-lepton (${p_{\top}(\tau)})$. Contrary to the case of quark 
fragmentation, the reconstructed $\tau$-jet momentum may be larger than the
 momentum of parent $\tau$-lepton (see Fig.~\ref{fig7}) due to the detector
 resolution and the $\tau$-jet reconstruction  algorithm.

We use the $ D_{\tau \, \to \, j} (z)$ parameterisation~(\ref{dz})
for the fit of the corresponding $z$-distributions for all considered
values of charged Higgs mass, namely for $m_H = 200,$ 250, 300, and 400~GeV
(see Fig.~\ref{fig7} and Table~\ref{tab2}).
We do not need achieve a good fit in the whole region of $z$.
We are interested in the values of $z$ close to unity, because this region 
corresponds to maximal values of $M_{\top}(j,\ETM)$ close to $M_H$.
Therefore, in what follows we use the set of parameters for 
$D_{\tau \to j}(z)$ from~(\ref{dz}) given below:
\begin{equation}
\alpha = 6.5, \quad \lambda = 3.5, \quad {\rm and} \,\,\, z_0 = 1.22
\end{equation}


Then, the  distribution of $M_{\top}(j, \ETM)$ could be evaluated
by convoluting the $m_{\bot}(\tau \nu)$-distribution
of the $\tau$-lepton and neutrino
 ($\propto 1/\sqrt{M_H^2 - m_{\bot}(\tau \nu)}$) with
$ D_{\tau \, \to \, j} (z)$-fragmentation function~(\ref{dz}):
\begin{equation}
\frac{dN}{d \, M_{\top}(j, \ETM)} = F_S(M_{\top}, M_f) \equiv
F_0 \, \int^{z_0}_{M_{\top}/M_f} 
\frac{M_{\top}}{M_f} \, \frac{z^{\alpha-1} (z_0-z)^{\lambda}}
{\sqrt{z^2 - M^2_{\top}/M_f^2  }} dz,
 \label{mjmis} 
\end{equation}
where $F_0$ is the normalisation factor and $M_f$ is the mass of the charged
Higgs boson to be determined from the fit.

We perform the fit of the $M_{\top}(j, \ETM)$-distribution for pure signal
events by means of this parameterisation $F_S(M_{\top}, M_f)$. 
The results of the fit are shown in Fig.~\ref{fig8}.
One can see that the proposed parameterisation~(\ref{mjmis}) provides not only 
a rather good description of the shape of the $M_{\top}(j, \ETM)$-distribution,
but also makes possible the determination of the fitted parameter $M_{f}$, 
which is very close to the input mass of charged Higgs boson.

\begin{figure}[b]
\begin{center}
\epsfig{file=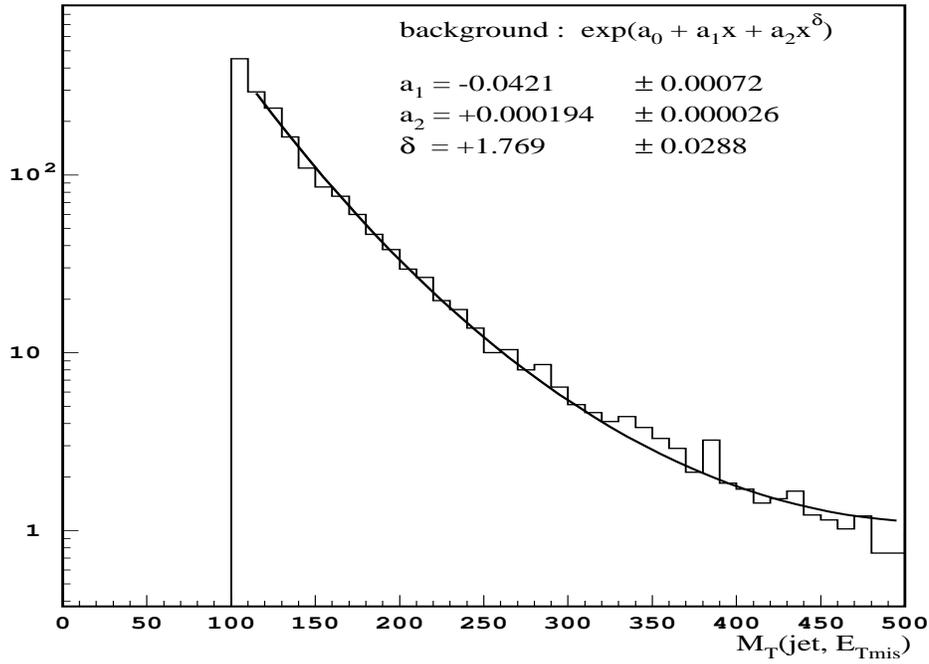,width=14cm,height=10cm,clip=} 
\caption{ The result of the fit of $M_{\top}(j, \ETM)$-distribution for 
the background events. }  \label{fig9} 
\end{center}
\end{figure}

For the $M_{\top}(j, \ETM)$-distribution of the background events we use a 
simple exponential parameterisation as follows:
\begin{equation}
F_B(M_{\top}) = exp(a_0 + a_1 M_{\top} + a_2 M_{\top}^{\delta}) \label{mexp}
\end{equation}
We get the following values of these parameters (see Fig.~\ref{fig9}):
\begin{equation}
\begin{array}{ll}
a_0 = 9.65 \pm 0.067, & a_1 = -0.0421 \pm 0.00072, \\ 
 a_2 = 0.000194 \pm 0.0000261, & \delta = 1.769 \pm 0.0288
\end{array} 
 \label{parexp}
\end{equation}

\section{\bf Signal visibility and measurement of charged Higgs parameters }

In the fitting procedure of the $M_{\top}(j, \ETM)$-distribution of the joint
 sample
of signal and background events we fix all parameters in the 
parameterisations $F_S(M_{\top}, M_f)$ and $F_B(M_{\top})$ except the
corresponding normalisations ($F_0$ in $F_S$ and $a_0$ in $F_B$). The
mass  of the \HP-boson ($M_f$ in $F_S(M_{\top}, M_f)$) is also left
as a free parameter to be determined by the fit.

\begin{figure}[htb]
\begin{center}
\epsfig{file=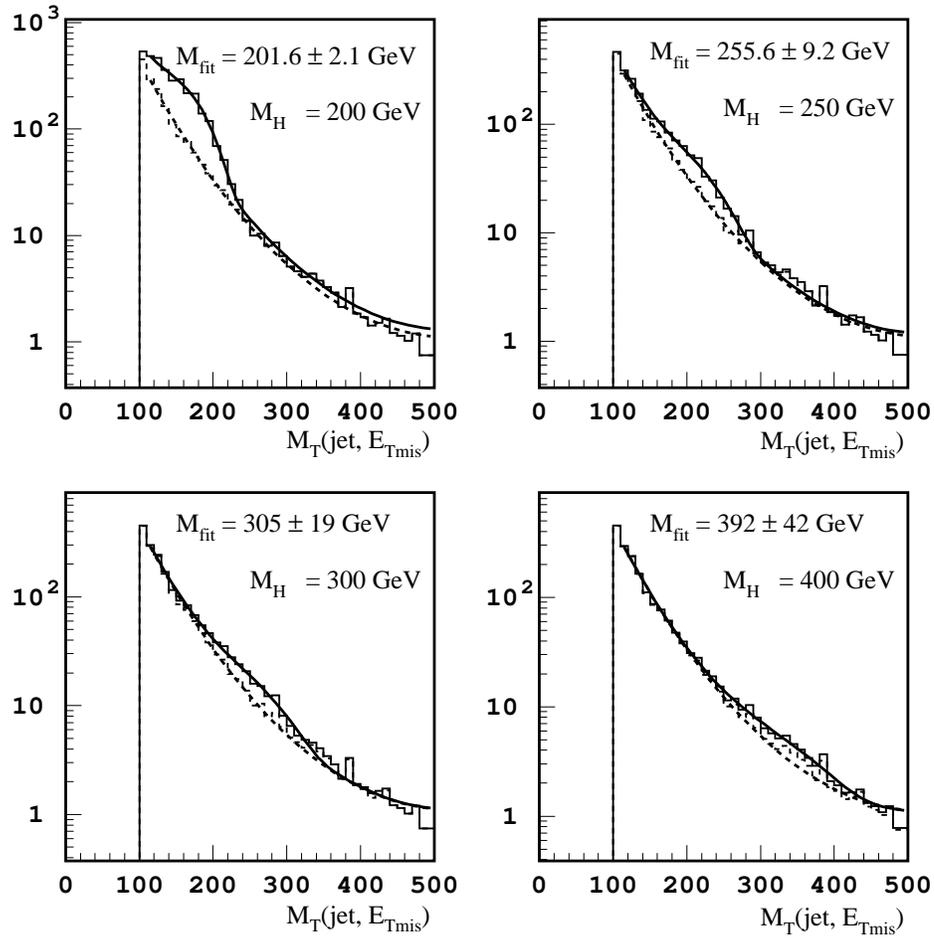,width=14cm,clip=}        
\caption{ The results of the fit of joint signal and background
$M_{\top}(j, \ETM)$-distribution by the sum of $F_S$ and $F_B$ 
parameterisations. } \label{fig10} 
\end{center}
\end{figure}

The results of the fit are shown in Fig.~\ref{fig10} and Table~\ref{tab3}. 
One can see that the extracted charged Higgs masses ($M_f$) from the 
fitting procedure coincide with input values ($M_H$) within the errors.
Therefore, the proposed parameterisation of the 
$M_{\top}(j, \ETM)$-distributions for signal events provides a 
reasonable way for a determination of the charged Higgs mass.

Then, using the normalisation parameters $F_0$ and $a_0$ for the signal and
background events we may evaluate the corresponding number of events:
\begin{eqnarray*}
 N_S^{f} &=& \int_{M_{\top, min}} F_S(M_{\top}, M_f) d M_{\top}, \\
 N_B^{f} &=& \int_{M_{\top, min}} F_B(M_{\top}) d M_{\top}
\end{eqnarray*}
These number are given in Table~\ref{tab3}. One can see that the values of
$N_B^{f}$ and $N_S^{f}$ extracted from the fitting procedure are in a good
agreement with the expected (generated) numbers $N_B$ and $N_S$ (see
Table~\ref{tab1}).

As a criterion for detection of the signal we use a significance criterion 
as follows  (which corresponds to 99\%~CL):
\begin{equation}
\frac{N_S^{f} }{\sqrt{N_S^{f} + N_B^{f}}} \geq 3   \label{cl}
\end{equation}
Due to our cuts where we require $E_{\top}(j)(\ETM) \geq 50$~GeV 
we perform the fit of the $M_{\top}(j, \ETM)$-distributions of the signal and
background events (see Fig.~\ref{fig10} and Table~\ref{tab3}) for $M_{\top}$
values greater than $M_{\top, min} = 100$~GeV.

From Table~\ref{tab3} one can see that the significance of the fit for 
$M_H \geq 300$~GeV is too low. To increase the signal-to-background 
ratio we truncate the $M_{\top}$ distribution, i.e. we repeat the same fitting
procedure, but for new value of $M_{\top, min}$, which is equal to
one half of $M_f^{(1)}$:
\[
  M_{\top, min}^{(2)} = \frac{1}{2} M_f^{(1)},
\]
and determine the new value for $M_f^{(2)}$. This decreases
significantly the number of the background events. The results of this new
fitting procedure are given in Table~\ref{tab4}. One obtains
almost the same results for $M_f$, but with on increased significance 
(compare Table~\ref{tab3} and Table~\ref{tab4}).

As explained in Section~1, for $M_H \geq 200$~GeV the production cross section 
for \HP has the almost quadratic dependence on $\tan\beta$ 
(see~(\ref{pp2})). Comparing the number of signal events extracted 
from the fit ($N_S^{f}$)  with that expected ($N_S$) in the MSSM scenario,
  we could determine the 
$\tan\beta$ parameter by means of simple equation:
\begin{equation}
\tan\beta_{f} = 50 \sqrt{ \frac{N_S^{f}}{N_S(M_f, \tan\beta=50)}},
\end{equation}
where $N_S(M_f, \tan\beta=50)$ is the number of expected events with
\HP-boson, generated with $M_H = M_f$ and $\tan\beta = 50$
(our default parameters, see~(\ref{defpar})). 
The corresponding uncertainty is evaluated as follows:
\begin{equation}
\delta (\tan\beta_f) = \frac{\tan\beta_f}{2}\sqrt{\delta^2_N + \delta^2_M }
\end{equation}
where $\delta_N$ is the relative error due to $F_S$ parameterisation.
A second relative error ($\delta_M$) is due to variation of the cross
section under $M_H$ variation within its own errors:
\[
\delta_M = \frac{1}{2\sigma} |\sigma(M_f-\Delta M) - \sigma(M_f+\Delta M)|
\]

\begin{figure}[t]
\begin{center}
\epsfig{file=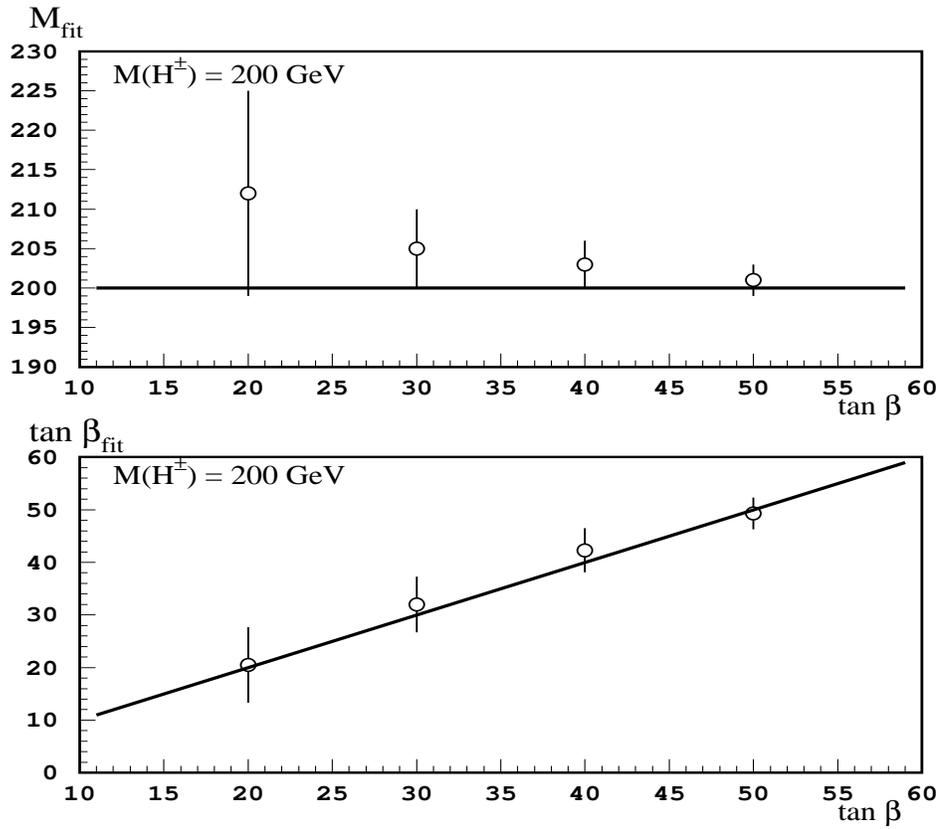,width=14cm,height=12cm,clip=} 
\caption{ The fitted values of \HP-boson mass and $\tan\beta$-parameter
for several values of the input $\tan\beta$. } \label{fig11} 
\end{center}
\end{figure}

The values of fitted $M_f$ for several  input $\tan\beta$ values are
given in Table~\ref{tab4} and Fig.~\ref{fig11}. Naturally, decreasing 
$\tan\beta$ leads to decreasing the number of signal events. As a result
the error in $M_f$ is increased. The corresponding extracted values of
$\tan\beta$ (for input $M_H = 200$~GeV) are given in the Table~\ref{tab5}
and Fig.~\ref{fig11}. 

\begin{figure}[b]
\begin{center}
\epsfig{file=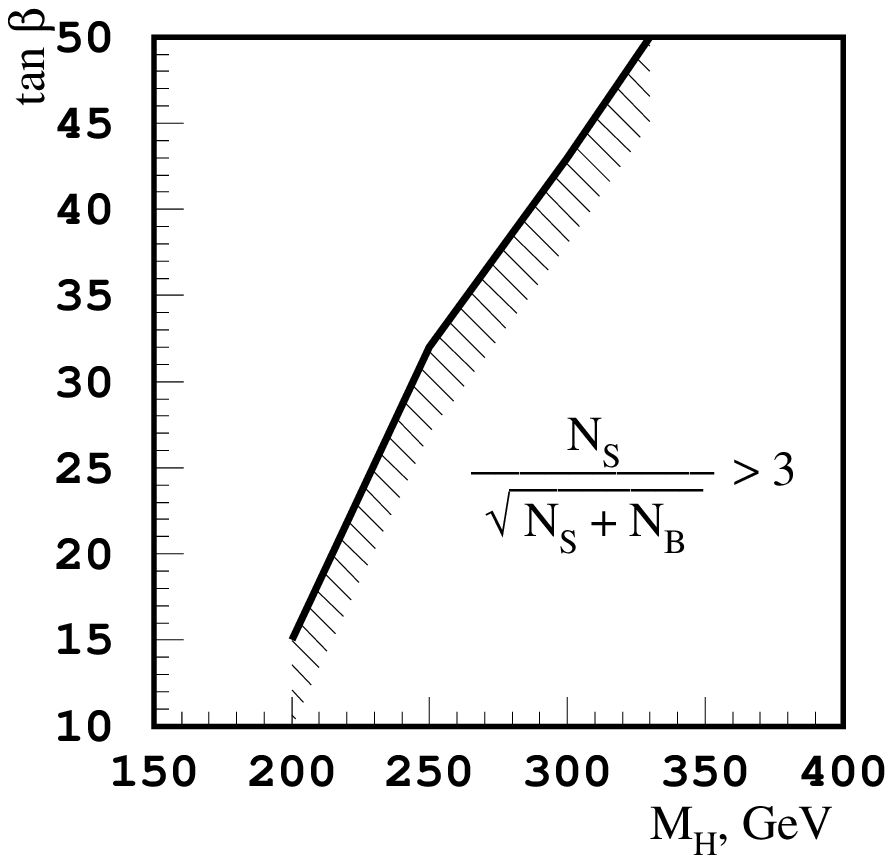,width=10cm,height=8cm,clip=} 
\caption{ The allowed region in ($M_H \, \times \, \tan\beta$) plot, which 
will be available for investigation in the of $s$-channel \HP-boson 
production (the left upper corner). } \label{fig12} 
\end{center}
\end{figure}

Using the criterion~(\ref{cl}) we evaluate the area in the
($M_H \, \times \, \tan\beta$)-MSSM plot which could be explored with
this process of $s$-channel \HP-boson production followed by a 
decay to $\tau^{\pm} \nu_{\tau}$ and a hadronic $\tau$ decays.
This region (the upper left corner) is shown in Fig.~\ref{fig12}.

\section {\bf Conclusion }

We investigate the CMS potential for study of \HP-boson production
via $s$-channel annihilation of light quarks. The study is made for large
value of $\tan\beta$ ($=20\div 50$) and relatively light charge Higgs 
boson ($M_H = 200 \div 400$~GeV) where the method is promising. 
Simple cuts are 
proposed for signal-to-background separation. After application of these cuts
a relatively large number of signal events ($N_S \sim 10^2 \div 10^3$)
may be expected. Therefore for this region of \HP-boson parameter space
 the study of \HP-boson production is possible with good a significance.

We find appropriate parameterisations for $M_{\top}(j, \ETM)$-distributions
of the signal and background events, which allow to determine the 
mass of charged Higgs, which is very close to the input values of $M_H$. 
Comparing the number of signal events evaluated from the fit,
with the expected ones in the framework of the MSSM, we can determine 
$\tan\beta$ with a 
reasonable accuracy. Using a standard significance criterion we determine 
region in the ($M_H \times \tan\beta$) parameter plot, where this
method could be applied. 

Finally, we conclude that the proposed subprocess of $s$-channel 
for the \HP-boson production does provide a good possibility for detecting
a relatively light charged Higgs boson and the measurement of its
mass and $\tan\beta$, 

\section* {\bf Acknowledgements}

The author thanks D.~Denegri for proposing to study this topic. 
Many thanks also should be given to  S.~Abdullin, H.-J.~He, R.~Kinnunen, 
A.~Miagkov, 
A.~Nikitenko, V.~Obraztsov,  and N.~Stepanov for  fruitful  discussions.


\newpage

\begin{table}[t]
\caption{The total number of the signal and background events after 
application of all cuts. Signal events are generated for 
$\tan\beta = 50$ and four values
of \HP-boson mass. 
The integrated luminosity is $L_{int} = 30$~fb$^{-1}$. } \label{tab1} 
\vspace{0.2cm}
\begin{center}
\begin{tabular}{|l|r|l|l|l||l|} \hline
 $M_H$, input & $200$  & $250$  & $300$  & $400$ & bkg  \\ \hline
 $N_{ev}$     & $1627$ & $344$  & $129$  & $35$  & $1756$ \\ \hline
 $\frac{N_S}{\sqrt{N_S + N_B}}$ & $28$ & $7.5$ & $3.0$ & $0.83$ & \\ \hline
\end{tabular}
\end{center}
\end{table}

\begin{table}[t]
\caption{The parameters resulted from the fit to ``fragmentation''
function $D_{\tau \, \to \, j} (z)$ from~(\ref{dz}). The fit was performed for
four \HP-boson mass values and in the region 
of $0.65 < z < 1.2$ ($z = {p_{\top}(j)} /{p_{\top}(\tau)}$). }
\label{tab2} 
\vspace{0.2cm}
\begin{center}
\begin{tabular}{|l|l|l|l|r|} \hline
 $M_H$ (GeV) & $\alpha$ & $\lambda$ & $z_0$(fixed) & $\chi^2/N$  \\ \hline
200 & $6.9 \pm 0.4$ & $3.35 \pm 0.17$ & $1.22$ & $42. / 23$ \\ \hline
250 & $6.5 \pm 1.0$ & $3.09 \pm 0.32$ & $1.22$ & $3.2 / 23$ \\ \hline
300 & $6.6 \pm 1.8$ & $3.49 \pm 0.61$ & $1.22$ & $1.8 / 23$ \\ \hline
400 & $6.1 \pm 4.2$ & $3.26 \pm 1.44$ & $1.22$ & $0.23/ 23$ \\ \hline
\end{tabular}
\end{center}
\end{table}

\begin{table}[t]
\caption{Results of the fit of joint signal and background events
$M_{\top}(j, \ETM)$-distribution for $M_{\top} \geq 100$~GeV. } \label{tab3}
\vspace{0.2cm}
\begin{center}
\begin{tabular}{|l||r|r|r||c|} \hline
 $M_H$ & $M_f$~(GeV) & $N^{f}_B$ & $N^{f}_S$ & 
$\frac{N^{f}_S}{\sqrt{N^{f}_B + N^{f}_S}}$   \\ \hline
 $200$ & $202. \pm 2.1$ & $1694$ & $1444$ & $25.8$ \\ \hline 
 $250$ & $256. \pm 9.2$ & $1694$ & $ 250$ & $5.9$  \\ \hline
 $300$ & $305. \pm 20.$ & $1694$ & $ 115$ & $2.7$  \\ \hline 
 $400$ & $392. \pm 42.$ & $1694$ & $  41$ & $1.0$  \\ \hline 
\end{tabular}
\end{center}
\end{table}

\begin{table}[t]
\caption{Results of the same fit as in Table~\ref{tab3}, but for
$M_{\top} \geq M_f/2$, where $M_f$ is result of previous fit.} \label{tab4} 
\vspace{0.2cm}
\begin{center}
\begin{tabular}{|l|l||r|r|r||c|} \hline
 $M_H$  & $M_{\top, min}$ & $M_f$~(GeV) & $N^{f}_B$ & $N^{f}_S$ & 
$\frac{N^{f}_S}{\sqrt{N^{f}_B + N^{f}_S}}$ \\ \hline
 $200$ & $\geq 100$ & $202. \pm 2.1$ & $1694$ & $1444$ & $25.8$ \\ \hline 
 $250$ & $\geq 125$ & $256. \pm 9.4$ & $899 $ & $ 231$ & $6.9$  \\  \hline
 $300$ & $\geq 150$ & $300. \pm 19.$ & $506 $ & $  97$ & $4.0$  \\ \hline 
 $400$ & $\geq 200$ & $391. \pm 43.$ & $245 $ & $  32$ & $1.9$  \\ \hline 
\end{tabular}
\end{center}
\end{table}

\nopagebreak
\begin{table}[b]
\caption{Results of the fit of joint signal and background events
$M_{\top}(j, \ETM)$-distribution for $M_H = 200$~GeV and several input values 
of $\tan\beta$. } \label{tab5} 
\vspace{0.4cm}
\begin{center}
\begin{tabular}{|l||c|c|c|c|c|} \hline
 $\tan\beta$  & $50$ & $40$ & $30$ & $20$ & $15$ \\ \hline
$M_{f}$ & $201 \pm 2$ & $203 \pm 3$& $205 \pm 5$ &$212 \pm 13$ &$222 \pm 28$ 
         \\ \hline 
$\tan\beta_{f}$ & $48.3 \pm 2.6$ & $39.3 \pm 4.1$ & $31.3 \pm 5.2$ &
   $19.8 \pm 7.6$ & $16.2 \pm 10.6$ \\ \hline
\end{tabular}
\end{center}
\end{table}

\newpage
\clearpage
\pagebreak 
\newpage

\vspace*{2.truecm}

\vspace{3.truecm}

\newpage

\newpage

\begin{thebibliography}{**}

\bibitem{Nilles:1984ge}
H.P.~Nilles, {\em Phys. Rept.\/}  {\bf 110} (1984) 1; \\
H.E.~Haber and G.L.~Kane, {\em Phys. Rept.\/}  {\bf 117} (1985) 75; \\
J.F.~Gunion, H.E.~Haber, G.L.~Kane, and S.~Dawson in 
{\it The Higgs Hunters' Guide} (Addison-Wesley, reading, MA, 1990).

\bibitem{Holzner:2001tv}
A.G.~Holzner, {\it Searches for charged Higgs bosons at LEP},
XXXVI Rencontres de Moriond, Electroweak Interactions and Unified Theories, 
Les Arcs, France, March 2001, arXiv:hep-ex/0105045.

\bibitem{Abe:1997rk}
F.~Abe {\it et al.}  [CDF Collaboration], {\em Phys. Rev. Lett.\/} 
{\bf 79} (1997) 357 [arXiv:hep-ex/9704003]; \\
T.~Affolder {\it et al.}  [CDF Collaboration], {\em Phys. Rev.\/} 
{\bf D62} (2000) 012004 [arXiv:hep-ex/9912013]; \\
B.~Abbott {\it et al.}  [D0 Collaboration], {\em Phys. Rev. Lett.\/}  {\bf 82} 
(1999) 4975 [arXiv:hep-ex/9902028]; \\
V.M.~Abazov {\it et al.}  [D0 Collaboration], arXiv:hep-ex/0102039.

\bibitem{Mangano:1997md}
M.L.~Mangano and S.R.~Slabospitsky, {\em Phys. Lett.\/} {\bf B410} (1997) 299
[arXiv:hep-ph/9707248].

\bibitem{atlas}  K.A.~Assamagan, ATLAS Internal Note ATL-PHYS-99-013 (1999),
ATL-PHYS-99-025 (1999), ATL-PHYS-2000-031 (2000); \\
K.A.~Assamagan, Y.~Coadou and A.~Deandrea,
arXiv:hep-ph/0203121.


\bibitem{cms} R.~Kinnunen, CMS Internal Note CMS NOTE 2000/045 (2000).

\bibitem{Denegri:2001pn}
D.~Denegri {\it et al.},
``Summary of the CMS Discovery Potential for the MSSM SUSY Higgses,''
CMS-NOTE-2001-032 (2001), arXiv:hep-ph/0112045.

\bibitem{roy}
D.P.~Roy,
{\em Phys. Lett.\/} {\bf B459} (1999) 607 [arXiv:hep-ph/9905542]; \\
M.~Drees, M.~Guchait and D.P.~Roy,
{\em Phys. Lett.\/}  {\bf B471} (1999) 39 [arXiv:hep-ph/9909266]; \\
D.P.~Roy, arXiv:hep-ph/0102091.

\bibitem{tait} 
H.-J.~He and C.-P.~Yuan,
{\em Phys. Rev. Lett.\/}  {\bf 83} (1999) 28 [arXiv:hep-ph/9810367]; \\
T.M.~Tait, hep-ph/9907462; \\
T.M.~Tait and C.-P.~Yuan,
{\em Phys. Rev.\/} {\bf D63} (2001) 014018 [arXiv:hep-ph/0007298].

\bibitem{Foursa:2000yu}
M.V.~Foursa, D.A.~Murashov, and S.R.~Slabospitsky, arXiv:hep-ph/0008198.

\bibitem{Diaz-Cruz:2001gf}
J.L.~Diaz-Cruz, H.-J.~He and C.-P.~Yuan,
arXiv:hep-ph/0103178.

\bibitem{Lai:1999wy}
  H.L.~Lai {\it et al.}   [CTEQ],  arXiv:hep-ph/9903282.

\bibitem{toprex} 
S.R.~Slabospitsky and L.~Sonnenschein,
``TopReX generator (version 3.25): Short manual,''
arXiv:hep-ph/0201292.

\bibitem{pythia}
T.~Sj\"ostrand and M.~Bengtsson, {\em Comput. Phys. Commun.\/} {\bf 43}, 367
(1987); T. Sj\"ostrand, {\it PYTHIA 5.7}, Comput. Phys. Comm. {\bf
82}, 74 (1994)

\bibitem{cmsjet}
S. Abdullin, A. Khanov, N. Stepanov, {\it CMSJET}, CMS TN/94-180 (1994).

\bibitem{Groom:2000in}
D.E.~Groom {\it et al.}  [Particle Data Group Collaboration],
 {\em Eur. Phys. J.\/} {\bf C15} (2000) 1.

\bibitem{Balazs:1998sb}
C.~Balazs, H.-J.~He and C.-P.~Yuan,
{\em Phys. Rev.\/} {\bf D60} (1999) 114001
[arXiv:hep-ph/9812263].


\bibitem{boos} E.E.~Boos, L.V.~Dudko, and V.I.~Savrin,
 CMS Internal Note CMS NOTE 2000/065 (2000).

\bibitem{klp} 
V.G.~Kartvelishvili, A.K.~Likhoded and V.A.~Petrov,
{\em Phys. Lett.\/} {\bf B78}, 615 (1978).

\end{thebibliography}
\end{document}